\documentclass[aps,prl,twocolumn,superscriptaddress,nofootinbib,floatfix]{revtex4-2}

\usepackage[T1]{fontenc}
\usepackage[utf8]{inputenc}
\usepackage{amsmath,amssymb,bm,mathtools}
\usepackage{graphicx}
\usepackage{xcolor}
\usepackage{hyperref}
\usepackage{microtype}
\usepackage[percent]{overpic}

\hypersetup{colorlinks=true,linkcolor=blue,citecolor=blue,urlcolor=blue}

\newcommand{\dd}{\mathrm{d}}
\newcommand{\ee}{\mathrm{e}}
\newcommand{\Calpha}{C_{\alpha}}
\newcommand{\fr}{f_r}
\newcommand{\deff}{d_{\mathrm{eff}}}

\begin{document}

\title{{Confinement Alters Energy Relaxation in Subrecoil Laser Cooling}}

\author{Daiki Kanbe}
\affiliation{Department of Physics, Tokyo University of Science, Noda, Chiba 278-8510, Japan}

\author{Eli Barkai}
\affiliation{Department of Physics, Institute of Nanotechnology and Advanced Materials, Bar-Ilan University, Ramat-Gan, Israel}

\author{Takuma Akimoto}
\affiliation{Department of Physics, Tokyo University of Science, Noda, Chiba 278-8510, Japan}

\date{\today}

\begin{abstract}
Controlling atomic motion through laser cooling requires both reducing kinetic energy and confining atoms to a finite spatial region. Here, we investigate how harmonic confinement modifies the long-time energy relaxation in subrecoil laser cooling. Using a stochastic scattering-rate model, numerical simulations, and an analytically tractable renewal approximation, we show that confinement qualitatively changes the asymptotic relaxation law. For a scattering rate that vanishes at low momentum as $R(p)\propto |p|^\alpha$, the change arises because confinement reshapes the phase-space geometry of rare, long trapping events. In free space, such events require only small momentum. Under harmonic confinement, however, position and momentum are coupled along phase-space orbits, restricting long-lived low-scattering trajectories to the vicinity of the phase-space origin. This geometric constraint modifies the trapping-time statistics, produces two dynamical crossovers, and leads to an optimal scattering-rate exponent of $\alpha=4$ for the fastest asymptotic decay of the mean energy.
Our results demonstrate that confinement changes the asymptotic energy-relaxation dynamics even when the local scattering law remains unchanged.
\end{abstract}

\maketitle

\textit{Introduction.---}
Laser cooling is a fundamental technique for controlling the motion of atoms at ultralow temperatures~\cite{Hansch1975,Wineland1978,Dalibard1989,CohenTannoudjiPhillips1990,Chu1998,CohenTannoudji1998,Phillips1998,Metcalf1999}.
Beyond its importance for precision measurements and quantum technologies~\cite{Cronin2009,Katori2011,DereviankoKatori2011,Ludlow2015,CornellWieman2002,Anderson1995,Davis1995,Bloch2012,Urvoy2019}, laser-cooled atoms provide a versatile platform for studying nonequilibrium statistical mechanics~\cite{BardouBook2002,BertinBardou2008,Afek2023}.
Unlike ordinary thermal systems that relax toward Boltzmann--Gibbs equilibrium, laser-cooled atoms are continuously driven by stochastic photon-scattering processes and can display broad non-Gaussian fluctuations and L\'evy-type statistics~\cite{Marksteiner1996,Sagi2012,KesslerBarkai2012,BarkaiAghionKessler2014,BarkaiRadonsAkimoto2021,BarkaiRadonsAkimoto2022JCP}.

A paradigmatic example is subrecoil laser cooling~\cite{Aspect1988,Aspect1989,KasevichChu1992,Bardou1994,Davidson1994,Lawall1994,Reichel1995,Esslinger1996,Saubamea1999,Boyer2004,Zhang2025Lithium}.
In this regime, atoms with sufficiently small momentum become increasingly unlikely to undergo further photon-scattering events.
At a coarse-grained level, the dynamics can be described as a heterogeneous random walk in momentum space: after each scattering event the atomic momentum $p$ changes, while the trapping time until the next event depends strongly on the current momentum~\cite{BardouBook2002,BertinBardou2008,AkimotoBarkaiRadons2022,ShirakiBarkaiAkimoto2026}.

For small momentum, the scattering rate typically scales as
$R(p)\propto |p|^\alpha$.
Consequently, rare exceptionally long trapping intervals occur close to
$p=0$, allowing the momentum distribution to become increasingly
concentrated around zero.
In free space, where the momentum remains constant between scattering
events, this mechanism gives rise to a fat-tailed trapping-time
distribution $\psi(\tau)\propto \tau^{-(1+1/\alpha)}$, which underlies the emergence of L\'evy statistics in subrecoil cooling~\cite{BardouBook2002,BertinBardou2008,AkimotoBarkaiRadons2022}.

The exponent $\alpha$ is protocol dependent rather than a purely abstract parameter.
Velocity-selective coherent population trapping realizes $\alpha=2$~\cite{Aspect1988}, while Raman-cooling protocols have realized $\alpha=2$ with square pulses and $\alpha=4$ with Blackman pulses~\cite{Reichel1995,ShirakiBarkaiAkimoto2026}.
More generally, different cooling protocols and pulse shapes can in principle realize distinct low-momentum scattering laws~\cite{Aspect1988,KasevichChu1992,Reichel1995,ShirakiBarkaiAkimoto2026}.
The long-time relaxation dynamics of subrecoil cooling in free space have been extensively studied~\cite{BardouBook2002,BertinBardou2008,AkimotoBarkaiRadons2022,ShirakiBarkaiAkimoto2026}.
The same rare-event mechanism is also connected with anomalous diffusion, infinite invariant densities, and weak ergodicity breaking~
\cite{BarkaiRadonsAkimoto2021,BarkaiRadonsAkimoto2022JCP,
AkimotoBarkaiRadons2022,ShirakiBarkaiAkimoto2026,Afek2023}.
Related intermittency mechanisms also arise in Pomeau--Manneville-type
dynamics and nonlinear oscillators~
\cite{MeyerKantz2017,MeyerAdlakhaKantzBassler2018}.

In an unconfined system, however, suppressing the momentum does not itself prevent atoms from spreading in real space.
Although this spreading can become slow at low energies, long-time experiments generally benefit from spatial trapping, which prevents atomic escape and enables long interrogation and manipulation times~\cite{Katori2011,DereviankoKatori2011,Ludlow2015}.
Subrecoil cooling has also been proposed for optically trapped atoms, where discrete motional states can be selectively protected by coherent population trapping~\cite{Lee2022MSCPT}.
Spatial confinement is therefore not merely an auxiliary ingredient: it can be essential for retaining the atoms during the long times required to exploit subrecoil cooling.

Despite the practical importance of confinement, most theoretical descriptions of subrecoil cooling have focused on unconfined motion~\cite{BardouBook2002,BertinBardou2008,BarkaiRadonsAkimoto2021,AkimotoBarkaiRadons2022,ShirakiBarkaiAkimoto2026}.
There, momentum remains constant between scattering events and directly determines the trapping time.
Under harmonic confinement, position and momentum are continuously interconverted.
Small instantaneous momentum alone therefore no longer guarantees a long trapping interval; long-lived low-energy trajectories must instead remain close to the origin of phase space.
Harmonic confinement thus provides a minimal setting in which spatial localization can alter the statistical mechanism of cooling~\cite{Lee1996DipoleTrap,Dechant2015,Dechant2016}.
This raises a more general question: how does harmonic confinement modify the statistics of rare trapping events and the resulting long-time energy relaxation, even when the underlying momentum-dependent scattering law remains unchanged?

In this Letter, we show that harmonic confinement changes the asymptotic mean-energy relaxation law by reshaping the phase-space geometry of rare trapping events.
The resulting modification of the trapping-time statistics produces two dynamical crossovers at $\alpha=2$ and $\alpha=4$, compared with $\alpha=1$ and $\alpha=3$ in free space~\cite{AkimotoBarkaiRadons2022}.
Our results demonstrate that the asymptotic relaxation dynamics depend not only on the low-momentum scaling of the scattering rate but also on the phase-space geometry imposed by confinement.

\textit{Model.---}
We consider a particle of unit mass confined by a one-dimensional harmonic potential.
{Between photon-scattering events,}
\begin{equation}
\dot{x}=p,\qquad \dot{p}=-\omega^2x,
\label{eq:ho}
\end{equation}
where $\omega$ is the trap frequency.
Photon-scattering events form a stochastic point process with instantaneous rate
\begin{equation}
R(p)=\frac{1}{\tau_0}\left|\frac{p}{p_0}\right|^\alpha.
\label{eq:rate}
\end{equation}
Thus an event occurs during $[t,t+\dd t]$ with probability $R[p(t)]\dd t+o(\dd t)$.
We use dimensionless variables, measuring time, momentum, and position in units of $\tau_0$, $p_0$, and $p_0\tau_0$, respectively, and write the dimensionless $\omega\tau_0$ simply as $\omega$.
{Hence $R(p)=|p|^\alpha$ below.}

We primarily study the continuous-position (CP) model, which is the direct confined extension of the exponential model.
At a scattering event occurring at time $t_s$, the position remains continuous, $x(t_s^+)=x(t_s^-)$, whereas the momentum is independently renewed from a symmetric density $f(p)$.
We assume only that $f(p)$ is smooth and finite near $p=0$.
The continuity of $x$ makes successive post-scattering states correlated even though the renewed momentum itself is independent.

For analytical purposes we therefore also introduce an auxiliary
reset-position (RP) model, in which both $x$ and $p$ are independently
renewed after every scattering event. This construction is related in spirit
to the broader framework of stochastic resetting~
\cite{EvansMajumdar2011,EvansMajumdarSchehr2020,PalKusmierzReuveni2019,BodrovaSokolov2020},
including generalized schemes in which the reset state is drawn from a distribution rather than fixed at a single point~
\cite{EvansMajumdar2011Optimal,GuptaJayannavar2022,Olsen2023}.

In the RP model used below, both reset variables are taken to be uniformly distributed on $[-1,1]$.
The asymptotic theory, however, requires only that the resulting post-scattering phase-space density be smooth and finite near the origin.
The deterministic harmonic motion and the momentum-dependent scattering rate are otherwise identical.
This distinction will be important below: the RP dynamics admits a closed renewal description, whereas the CP model is used as the physically motivated dynamics against which the analytical prediction is tested.

Since position and momentum are continuously interconverted by harmonic motion,
the relevant slow variable is no longer the instantaneous momentum but the
phase-space radius, which is directly related to the harmonic energy,
\begin{equation}
r=\sqrt{p^2+\omega^2x^2}=\sqrt{2E},
\label{eq:rE}
\end{equation}
where $E=r^2/2$ is the harmonic energy.
For a post-scattering state $(x_J,p_J)$, the subsequent motion is
\begin{align}
x(\tau)&=x_J\cos\omega\tau+\frac{p_J}{\omega}\sin\omega\tau,\\
p(\tau)&=p_J\cos\omega\tau-\omega x_J\sin\omega\tau.
\end{align}
Thus $r$ is exactly conserved between scattering events, while a scattering event changes the orbit by renewing the momentum and, in the RP model only, the position.
This conservation makes $r$ the natural variable for describing rare low-energy intervals under confinement.

\textit{Numerical implementation.---}
Between scattering events, the particle follows the deterministic harmonic trajectory, while the scattering rate $R[p(t)]=|p(t)|^\alpha$ varies continuously along the orbit.
We exploit this deterministic motion to simulate the corresponding nonhomogeneous Poisson process using the standard thinning method~\cite{LewisShedler1979,Ogata1981,TokarDreysse2025}.
Along an orbit of radius $r$, the momentum satisfies $|p(t)|\le r$, and hence the scattering rate is bounded by
$R[p(t)]\le r^\alpha$.
We therefore define the maximum scattering rate along the orbit as
$\Lambda=r^\alpha$.
Candidate interarrival times are sampled as
$\Delta t_c=-\ln U/\Lambda$, where $U$ is uniform on $(0,1)$.
At each candidate time $t_c$, the harmonic trajectory is advanced to $t_c$, and the candidate is accepted with probability
\begin{equation}
P_{\rm acc}(t_c)=\frac{R[p(t_c)]}{\Lambda}
=\left|\frac{p(t_c)}{r}\right|^\alpha.
\label{eq:pacc}
\end{equation}
Although the acceptance probability is evaluated from the instantaneous momentum at the candidate time, the sequence of candidate events and rejections incorporates the time dependence of the rate along the trajectory.
As a result, the survival probability up to time $\tau$ is
$S(\tau)=\exp[-\int_0^\tau R[p(t)]\,dt]$,
so that the accepted events are distributed according to the original time-dependent scattering rate.

For the free-space comparison, the harmonic force is set to zero and the waiting time is sampled directly from an exponential distribution with rate $|p|^\alpha$, using the same post-scattering momentum distribution.
In the simulations, the renewed momentum is drawn uniformly from $[-1,1]$ in dimensionless units.

\begin{figure}[t]
\centering
\begin{overpic}[width=0.90\linewidth]{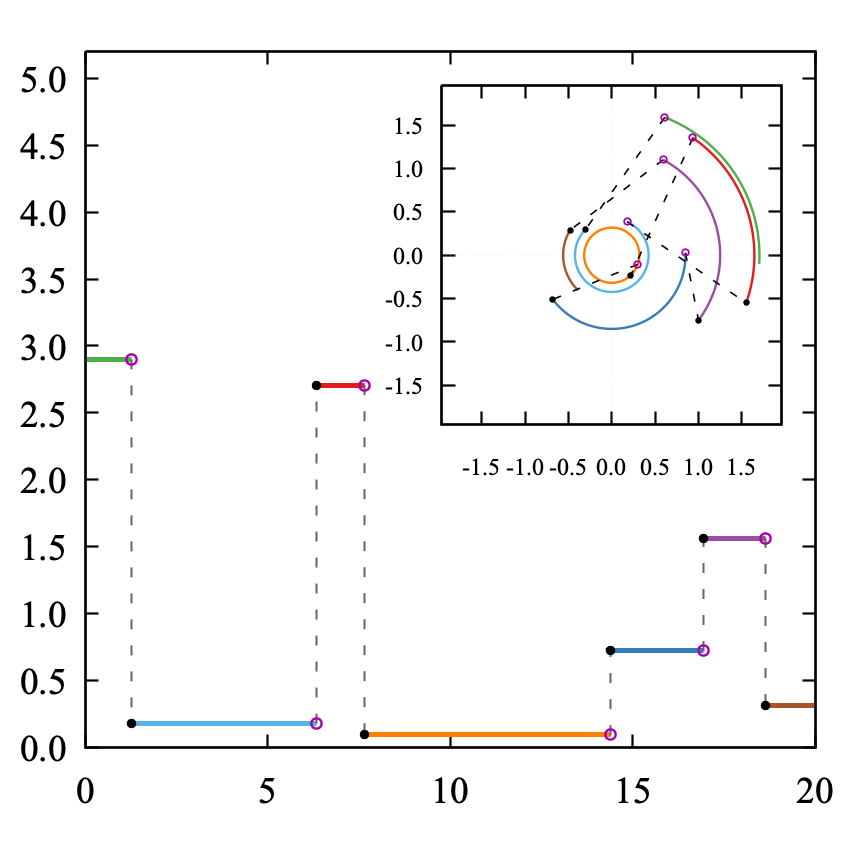}
\put(50,-2){\Large $t$}
\put(-8,47){\rotatebox{90}{\Large $r(t)$}}
\put(70,40){\large $p$}
\put(43,67){\rotatebox{90}{\large $\omega x$}}
\end{overpic}
\caption{Time evolution of the orbit radius $r(t)$ for $\omega=1$ and $\alpha=2$. Between scattering events $r$ is conserved, whereas scattering changes the orbit. Inset: corresponding trajectory in the $(\omega x,p)$ plane.}
\label{fig:rate_phase_relation}
\end{figure}

\textit{Energy relaxation.---}
We characterize the cooling dynamics by
\begin{equation}
\langle E(t)\rangle\propto t^{-\lambda_E},
\label{eq:energy_decay}
\end{equation}
where $\lambda_E$ is the asymptotic mean-energy relaxation exponent.
A larger $\lambda_E$ means faster algebraic decay; it does not imply a lower final temperature.
Different cooling measures need not have the same optimum: in the free-space exponential model the kinetic-energy decay is fastest at $\alpha=3$, whereas a criterion based on the half-width of the cooled momentum peak gives $\alpha=1$~\cite{AkimotoBarkaiRadons2022}.
We therefore use $\lambda_E$ here as a measure of asymptotic energy-relaxation efficiency, rather than as a measure of the ultimate cooling temperature.

\begin{figure}[t]
\centering
\begin{overpic}[width=0.96\linewidth]{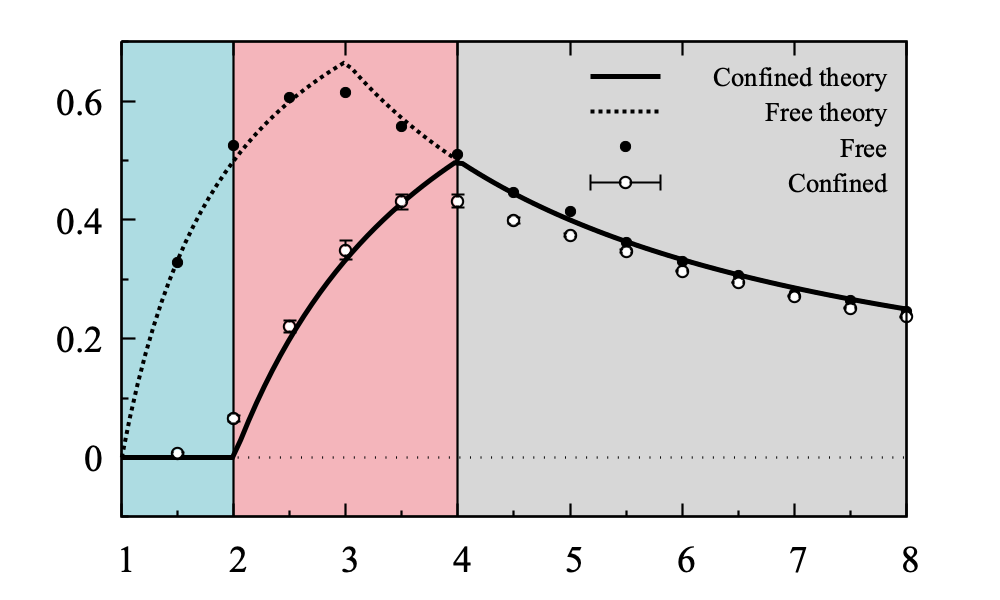}
\put(0,32){\makebox(0,0){\rotatebox[origin=c]{90}{\Large $\lambda_E$}}}
\put(50,0){\makebox(0,0){\Large $\alpha$}}
\end{overpic}
\caption{Energy-relaxation exponent $\lambda_E$ versus $\alpha$.
Filled circles: free-space simulations;
open circles: confined CP simulations.
The vertical error bars on the open circles indicate
the formal standard errors of the fitted exponents.
{Solid and dotted lines}:
confined RP theory and free-space theory, respectively.
Logarithmic corrections produce slow finite-time convergence
near $\alpha=2$ and $4$.
}
\label{fig:energy_relaxation_exponent}
\end{figure}

The analytical RP renewal theory derived below predicts
\begin{equation}
\lambda_E(\alpha)=
\begin{cases}
0, & 0<\alpha<2,\\[2pt]
1-2/\alpha, & 2<\alpha<4,\\[2pt]
2/\alpha, & \alpha>4,
\end{cases}
\label{eq:lambdaE}
\end{equation}
with logarithmic corrections at the marginal cases $\alpha=2$ and $4$.
The two crossover points have distinct physical origins, as explained below.
Figure~\ref{fig:energy_relaxation_exponent} compares Eq.~\eqref{eq:lambdaE} with simulations of the physical CP model.
The numerical exponents are obtained by least-squares fits of
$\ln\langle E(t)\rangle$ versus $\ln t$ in the long-time regime;
details of the fitting procedure and uncertainty estimation are given
in the Supplemental Material.
The numerical exponents follow the same three-regime structure and agree with the RP prediction away from the crossover points.
Close to $\alpha=2$ and $4$, convergence is especially slow because the corresponding asymptotic laws contain logarithmic corrections.
The exponent increases for $2<\alpha<4$ and decreases for $\alpha>4$, so its maximum occurs at $\alpha=4$.

For comparison, free-space subrecoil cooling has crossovers at $\alpha=1$ and $3$ and the mean-energy relaxation exponent is maximal at $\alpha=3$~\cite{BardouBook2002,AkimotoBarkaiRadons2022,ShirakiBarkaiAkimoto2026}.
Harmonic confinement therefore changes both crossover points and shifts the maximum from $3$ to $4$.
In particular, confinement reduces the asymptotic mean-energy relaxation exponent for $1<\alpha<4$.
This reveals a practical trade-off: confinement prevents the spatial escape of atoms, but it can reduce the asymptotic rate of energy relaxation.

\textit{Trapping-time statistics.---}
The origin of this change can be seen by first examining a single interval between successive scattering events.
Importantly, this conditional interval dynamics is the same for CP and RP once the post-scattering state is specified.
Write that state as $p_J=r\cos\phi$ and $\omega x_J=r\sin\phi$.
During the subsequent harmonic motion,
\begin{equation}
p(t)=r\cos(\omega t+\phi),
\end{equation}
so the instantaneous rate is $r^\alpha|\cos(\omega t+\phi)|^\alpha$.
The scattering rate integrated along the trajectory up to time $\tau$ is
\begin{equation}
H(\tau|r,\phi)=r^\alpha\int_0^\tau |\cos(\omega u+\phi)|^\alpha\dd u,
\end{equation}
and the corresponding survival probability is
\begin{equation}
S(\tau|r,\phi)=\ee^{-H(\tau|r,\phi)}.
\label{eq:hazard}
\end{equation}
Differentiating the survival probability gives the exact conditional trapping-time PDF,
\begin{equation}
q(\tau|r,\phi)
=
r^\alpha|\cos(\omega\tau+\phi)|^\alpha
\exp[-H(\tau|r,\phi)].
\end{equation}
It is not exactly exponential because the instantaneous scattering rate oscillates periodically along the orbit.
This exact expression is independent of the post-scattering reset rule and therefore applies equally to CP and RP trajectories.

For the rare long intervals that control the asymptotic tail, the atom typically completes many harmonic oscillations before scattering, $\omega\tau\gg1$.
The periodic modulation can then be separated from the slow radial dependence.
Introducing the cycle average
\begin{equation}
\Calpha=
\frac{1}{2\pi}
\int_0^{2\pi}
|\cos\theta|^\alpha\dd\theta,
\label{eq:calpha}
\end{equation}
we obtain
$
H(\tau|r,\phi)=
\Calpha r^\alpha\tau+
O(r^\alpha/\omega)$ 
in the scaling limit $r\to0$ with $r^\alpha\tau=O(1)$.
Hence
\begin{equation}
q(\tau|r)
\simeq
k(r)\ee^{-k(r)\tau},
\qquad
k(r)=\Calpha r^\alpha.
\label{eq:qr}
\end{equation}
This single-interval result applies to both CP and RP dynamics; the reset rule enters through the distribution of radii immediately after scattering.

Let $\rho_J(x,p)$ be the post-scattering joint density.
Its radial marginal is
\begin{equation}
\fr(r)=
\frac{r}{\omega}
\int_0^{2\pi}
\rho_J
\left(
\frac{r}{\omega}\sin\varphi,
r\cos\varphi
\right)
\dd\varphi.
\label{eq:fr}
\end{equation}
For the RP model, $\rho_J$ is fixed directly by the independent reset distributions.
In the CP model, by contrast, the position marginal is generated dynamically because the position remains continuous through a scattering event.
The asymptotic tail depends only on the behavior of $\fr(r)$ near the phase-space origin, which we parameterize as
\begin{equation}
\fr(r)\sim Ar^{\deff-1}.
\end{equation}
This defines an effective dimension $\deff$ for the low-energy geometry.
If $\rho_J(x,p)$ is smooth and finite at the origin, independently of whether it factorizes, the probability contained in a small disk scales with its area and Eq.~\eqref{eq:fr} gives $\fr(r)\propto r$.
Hence $\deff=2$ for a regular confined phase-space density.
The conclusion is therefore geometric rather than RP-specific.
In free space the relevant slow variable is the one-dimensional momentum, so a regular post-scattering momentum density corresponds to $\deff=1$.
Here, $\deff$ should not be interpreted as the spatial dimension of the system: the physical motion remains one-dimensional.
Rather, it characterizes the effective low-energy phase-space geometry that rare long-lived trajectories must explore.
Physically, the change from $\deff=1$ to $\deff=2$ reflects that only the momentum must be small in free space, whereas under harmonic confinement a long-lived low-scattering orbit requires a small phase-space radius, so both $p$ and $\omega x$ must be small.

\begin{figure}[t]
\centering
\begin{overpic}[width=\linewidth]{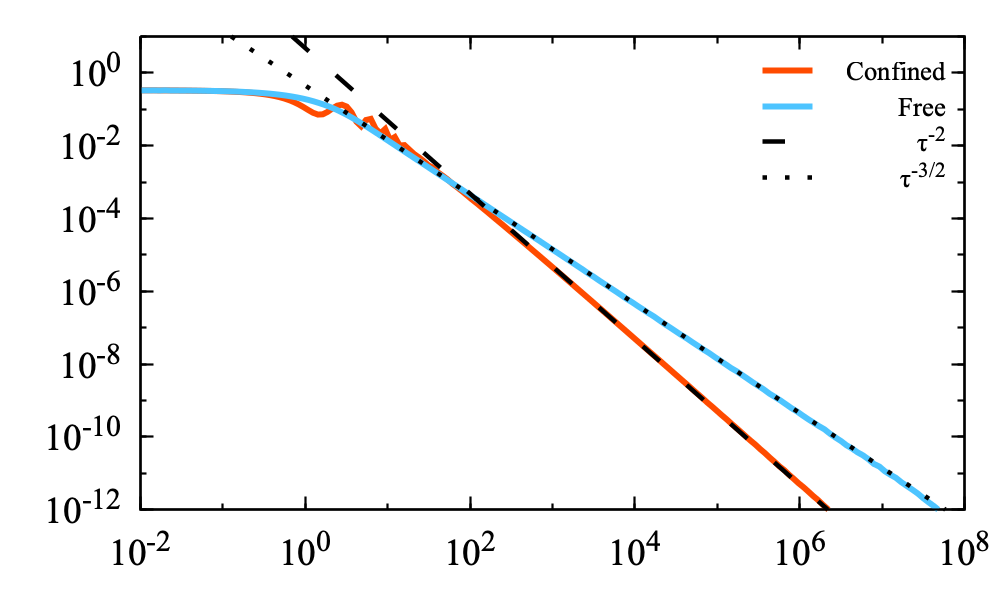}
\put(2,32){\makebox(0,0){\rotatebox[origin=c]{90}{\Large $\psi(\tau)$}}}
\put(54,0){\makebox(0,0){\Large $\tau$}}
\end{overpic}
\caption{Trapping-time PDF at $\alpha=2$. The confined CP model (orange) follows the predicted $\tau^{-2}$ tail, while the free-space model (blue) follows $\tau^{-3/2}$.}
\label{fig:trapping_time_pdf}
\end{figure}

Averaging Eq.~\eqref{eq:qr} over post-scattering radii gives
\begin{equation}
\psi(\tau)
\simeq
\int_0^\infty
\fr(r)\Calpha r^\alpha
\ee^{-\Calpha r^\alpha\tau}
\dd r
\propto
\tau^{-(1+\deff/\alpha)}.
\label{eq:psi_tail}
\end{equation}
The large-$\tau$ integral is dominated by radii
$r\sim(\Calpha\tau)^{-1/\alpha}$,
making the connection between rare long intervals and the small-$r$ geometry explicit.
Harmonic confinement therefore changes the free-space tail $\tau^{-(1+1/\alpha)}$ into $\tau^{-(1+2/\alpha)}$.
Figure~\ref{fig:trapping_time_pdf} illustrates this directly at $\alpha=2$: the confined CP simulation follows $\psi(\tau)\propto\tau^{-2}$, whereas the free-space model follows $\tau^{-3/2}$.
The agreement supports the assumption that the dynamically generated CP post-scattering density remains regular near the phase-space origin.

\textit{Origin of the two crossovers.---}
The two crossover points in Eq.~\eqref{eq:lambdaE} have distinct physical origins.
The trapping-time tail immediately explains the first crossover.
According to Eq.~\eqref{eq:psi_tail},
the mean trapping time changes from finite to divergent at $\alpha=\deff$.
The same threshold is obtained from the formal stationary occupation weight: trajectories enter a radius interval with probability $\fr(r)\dd r$ and remain there for a mean time proportional to $1/k(r)$, giving
\begin{equation}
I_r(r)\propto
\frac{\fr(r)}{k(r)}
\propto
r^{\deff-1-\alpha}.
\end{equation}
This measure is normalizable near $r=0$ only when $\alpha<\deff$.
For $\alpha\geq\deff$, this non-normalizable occupation weight corresponds to a radial infinite invariant density in the RP model, linking the cooling regime to infinite ergodic theory~\cite{AkimotoBarkaiRadons2022,Afek2023}.
For the confined geometry $\deff=2$, the first crossover therefore occurs at $\alpha=2$, separating a stationary regime from the nonstationary cooling regime.

The second crossover has a different origin because it depends on the observable.
During one trapping interval the harmonic energy is constant.
A low-$r$ interval has characteristic duration $\tau(r)\propto r^{-\alpha}$ and energy $E(r)\propto r^2$, so the time-integrated energy accumulated during that interval scales as
$\mathcal I(r)=E(r)\tau(r)\propto r^{2-\alpha}$.

Multiplying this interval reward by the small-$r$ phase-space weight gives
\begin{equation}
\fr(r)E(r)\tau(r)\,\dd r
\propto
r^{\deff+1-\alpha}\dd r,
\label{eq:reward}
\end{equation}
which becomes nonintegrable at $\alpha=\deff+2$.
Hence the two crossover points are
\begin{equation}
\alpha=\deff,
\qquad
\alpha=\deff+2:
\end{equation}
$(1,3)$ in free space and $(2,4)$ under harmonic confinement.
This identifies the two transitions with two distinct integrability conditions: the first concerns the residence time itself, whereas the second concerns the energy-weighted residence time.

\textit{Montroll--Weiss description of energy relaxation.---}
To obtain the complete mean-energy law we now specialize to the RP model, where successive post-scattering states form a renewal process.
For convenience, we formulate the renewal description in terms of the phase-space radius ($r=\sqrt{2E}$). The long-time radius density has the Montroll--Weiss form
\begin{equation}
\widetilde{\rho}(r,s)
\simeq
\frac{\fr(r)}
{s+\Calpha r^\alpha}
\frac{1}
{1-\widetilde{\psi}(s)}.
\label{eq:mw}
\end{equation}
For $\alpha>\deff$, Eq.~\eqref{eq:psi_tail} gives
$1-\widetilde{\psi}(s)\sim s^{\deff/\alpha}$.
Since $E=r^2/2$,
\begin{equation}
\widetilde{\langle E\rangle}(s)
\simeq
\frac{1}
{2[1-\widetilde{\psi}(s)]}
\int_0^\infty
\frac{\fr(r)r^2}
{s+\Calpha r^\alpha}
\dd r.
\label{eq:energy_laplace}
\end{equation}
For $\deff=2$, $\fr(r)\sim Ar$ and the integrand behaves as $r^{3-\alpha}$ above the crossover scale
$r_*(s)=(s/\Calpha)^{1/\alpha}$.
For $2<\alpha<4$ this integral remains finite as $s\to0$, so the singular renewal factor alone gives
$\widetilde{\langle E\rangle}(s)\sim s^{-2/\alpha}$,
and hence $\langle E(t)\rangle\sim t^{-(1-2/\alpha)}$.
For $\alpha>4$, the integral itself diverges at the origin and is dynamically cut off at $r_*(s)$.
{The} additional factor
$r_*^{4-\alpha}\sim s^{(4-\alpha)/\alpha}$
changes the asymptotic law to
$\langle E(t)\rangle\sim t^{-2/\alpha}$.

Together with the stationary regime for $\alpha<2$, this reproduces Eq.~\eqref{eq:lambdaE}.
At the marginal points, the corresponding integrals acquire logarithmic corrections.
In particular, the leading asymptotic forms are
\begin{equation}
\langle E(t)\rangle\sim \frac{1}{\ln t},
\qquad \alpha=2,
\end{equation}
and
\begin{equation}
\langle E(t)\rangle\sim t^{-1/2}\ln t,
\qquad \alpha=4.
\end{equation}
These logarithmic factors lead to especially slow convergence near the two crossover points and account for the finite-time deviations visible in Fig.~\ref{fig:energy_relaxation_exponent}.

The analytical treatment can be carried out explicitly for the RP model. The physical CP model is not an exact renewal process because the position remains continuous across scattering events. Nevertheless, the CP simulations reproduce both the predicted energy-relaxation exponents and the representative trapping-time tail. This agreement indicates that the asymptotic scaling is controlled primarily by the low-energy phase-space geometry, characterized by \(d_{\rm eff}=2\) under harmonic confinement, rather than by the microscopic details of the reset protocol. 
Here, $d_{\rm eff}$ is not the spatial dimension, but characterizes the number of phase-space coordinates that must simultaneously become small for long-lived low-scattering trajectories.

\textit{Quantum cutoff.---}
The present theory is based on a classical or semiclassical phase-space description of the atomic motion, in which the mean energy decays algebraically toward zero.
In a fully quantum treatment, however, the harmonic oscillator has the finite ground-state energy $E_0=\hbar\omega/2$~\cite{Monroe1995,Hamann1998}.
The scaling laws derived here should therefore be interpreted as intermediate asymptotics valid while $\langle E(t)\rangle\gg E_0$.
At sufficiently long times, quantum effects are expected to regularize the classical low-energy singularity and modify the approach to the lowest motional states.
Accordingly, $\lambda_E$ characterizes the relaxation dynamics before the quantum cutoff becomes dominant, rather than the ultimate temperature itself.

\textit{Conclusion.---}
Our results show that harmonic confinement qualitatively changes the long-time dynamics of subrecoil laser cooling by reshaping rare low-energy trajectories.
Although the local scattering law \(R(p)\propto |p|^\alpha\) is unchanged, confinement modifies the trapping-time statistics and asymptotic energy relaxation, revealing a trade-off between atomic retention and cooling performance.
Agreement between the RP theory and CP simulations supports this geometric interpretation and shows that a protocol optimized in free space need not remain optimal under confinement.
More generally, our results show that the asymptotic relaxation of subrecoil cooling is controlled jointly by the low-momentum scattering law and the effective low-energy phase-space geometry.

\textit{Acknowledgments.---}
This research was supported by The Israel Science Foundation (grant No. 2311/25).

\bibliographystyle{apsrev4-2}
\bibliography{references}

\end{document}